# Vicinal silicon surfaces: from step density wave to faceting

F.Leroy*, P.Müller*, J.J.Métois, O.Pierre-Louis**

Centre de Recherche sur la Matière Condensée et les Nanosciences,
UPR CNRS associé aux universités Aix Marseille II et III
Campus de Luminy, case 913, F-13288 Marseille Cedex 9, France
* also Université Paul Cézanne-Aix Marseille III
* * CNRS/The Rudolf Peierls Centre for Theoretical Physics
1 Keble Road, Oxford OX1 3NP, U.K.

**Abstract**

This paper investigates faceting mechanisms induced by electromigration in the regime where atomic steps are transparent. For this purpose we study several vicinal orientations by means of *in-situ* (optical diffraction, electronic microscopy) as well as *ex-situ* (AFM, microprofilometry) visualization techniques. The data show that faceting proceeds in two stages. The first stage is short and leads to the appearance of a step density wave, with a wavelength roughly independent of the surface orientation. The second stage is much slower, and leads to the formation of a hill-and-valley structure, the period of which depends on the initial surface orientation. A simple continuum model enables us to point out why the wavelength of the step density wave does not depend on the microscale details of the surface. The final wavelength is controlled by the competition between elastic step-step interaction and facet edge energy cost. Finally, the surface stress angular dependence is shown to emerge as a coarsed-grained picture from the step model.

**Introduction**

Due to its scientific and technological interest, faceting of stepped surfaces has been a long standing subject of intensive research [1-25]. Indeed, from a fundamental viewpoint the underlying mechanisms are still matter of debate. Furthermore, facetted systems appear to be promising templates for the "bottom-up" design of nanostructures.

One of the most important mechanisms for faceting is current-induced step bunching. While the instability of the surface is driven by electromigration [5,13-16], the resulting pattern arises from the interplay between electromigration-induced mass transport and the minimization of the elastic energy variations resulting from the changes in the surface morphology. As we shall see in the following, step bunching also appear as a promising way to study fundamental aspects of step-step elastic interactions as well to control the surface morphology at the micro or nano scale. At the nanoscale, considerable amount of research has been devoted to the understanding of the role of surface steps in the morphological evolution of vicinal Si(111) surfaces during sublimation [1-25]. These phenomena depend both on temperature and on the direction of the heating current. Stoyanov [16] was the first to propose a step model based on the Burton-Cabrera-Frank (BCF) model [26], in which electromigration is introduced as a bias in the Brownian motion of the adatoms on the surface [14-16]. At the microscale, Marchenko [27,28] then Alerhand [29] proposed a simple theory, based on elastic minimisation, to explain the micrometric periods which appear by annealing unstable surfaces.

In this paper we study the influence of the surface orientation on the instability, as well as the link between nanoscopic and microscale models. To do so, we have performed a systematic study of the surface morphology from the first stages (where the vicinal surface is described as a step pattern) towards the microscale state (where the facetted surface is described as a hill-and-valley structure formed by microscale facets) for various vicinal orientations. In all cases the morphological evolution proceeds in two stages: a short one based on the formation of a step density wave (the period of which, roughly does not depend upon the surface orientation) followed by a much slower one, where periodic microscale facets form via a step bunching mechanism. The final faceting period seems to depend on the elastic properties of the so-formed microscale facets.

The paper is divided in four parts. In the first part, the first section is devoted to the description of the vicinal faces under study then to a description of the experimental procedure. The experimental results are reported in part II. In part III we analyse the final state (III.1) and the first stages of faceting (III.2). The last part (part IV) consists in a short conclusion.

## I/ Description of the samples and of the experiments.

**I.1/ Morphological and elastic description of the vicinal faces under study:**

Vicinal surfaces can be described as stairs-like surfaces, where monatomic steps separate microscopically flat terraces. Since the atoms belonging to the step edges have a different number of nearest neighbors than the atoms in the underlying bulk, steps give rise to a lattice distortion that mediates an elastic interaction between them. The elastic description of the steps depends upon the state of the surface (surface of a stress-free body or surface of a stressed body for example [30]) as well as upon its structure. As a preamble, we thus would like to provide the reader with a detailed description of the geometry and the elastic properties of the surfaces that will be analyzed in the subsequent sections.

As shown in Fig. 1a, the selected vicinal orientations -(118), (223) (443) and (105) surfaces - form a closed cycle in the stereographic projection. More precisely, in Fig. 1a, are shown: in red the normal to the selected vicinal faces, and in blue, the normal to the (001), (113), (111), (110) and (100) surfaces which appear on the Si equilibrium shape [31]. In figure 1b are also reported the morphologies of some crystal surfaces with zone axis $[\bar{1}10]$. We may see, e.g., that the (118) surface is a vicinal of the (001) surface and thus is constituted of (001) terraces separated by monoatomic steps forming (111) micro-facets while (223) and (443) surfaces are vicinals of the (111) surface and thus are composed of (111) terraces separated by monoatomic steps forming (001) micro-facets. Furthermore, notice that the (001) and (111) surfaces are flat at the atomic scale (F surfaces) while the (113) and (101) surfaces can be considered to be flat at the second neighbor scale (at least for the fcc model).

In such a terrace/step model, important differences exist between the different vicinal surfaces under study. Let us thus consider separately vicinals of Si(111) and vicinals of Si(001).

 (1) Since the (111) surface is isotropic, vicinals of Si(111) exhibit equivalent (111) terraces characterized by isotropic surface stresses (see figure 2a). In other words, the surface stress is a scalar. From an elastic viewpoint, steps, separating the (111) terraces, can be modeled by rows of elastic dipoles distributed along the step edge [30]. The elastic interaction between steps per unit length then scales as $\ell^{-2}$ where $\ell$ is the inter-step distance (see appendix A).

 (2) Si(001) is not an isotropic surface. Indeed, its number of dangling bonds is reduced by the formation of dimer pairs aligned along the $\langle \bar{1}10 \rangle$ direction [32]. Thus, due to the diamond structure of silicon, two neighboring terraces separated by an atomic step do not have the same surface termination: one terrace exhibits a (1x2) reconstructed surface with dimers parallel to the $[\bar{1}10]$ direction, while the other terrace exhibits a (2x1) reconstructed surface with dimers parallel to the $[110]$ direction. In other words, two neighboring terraces exhibit two equivalent surface reconstructions rotated by 90° with respect to each other. Since the surface stress component parallel to the dimer axis is more tensile than the surface stress component perpendicular to the dimer axis [29], the surface stress of the (001) terraces is a second rank tensor which reads $\begin{pmatrix} s_{xx} & 0 \\ 0 & s_{yy} \end{pmatrix}$ for one terrace and $\begin{pmatrix} s_{yy} & 0 \\ 0 & s_{xx} \end{pmatrix}$ for the other (when written in the $[110]$, $[\bar{1}10]$ surface axis). As a consequence, the elastic description of the vicinal surfaces of Si(001) depends upon the azimuthal disorientation angle. More precisely:

(i) Ideal vicinal surfaces with $[110]$ zone axis (case of (113) and (118) ideal surfaces) are formed by steps parallel to the $[110]$ direction, (see figure 2b) so that the surface stress difference $\pm|s_{xx}-s_{yy}|$ in the direction normal to the step gives birth to a net force across the step. The action of these steps on the underlying crystal can be modeled by a line of elastic monopoles perpendicular to the steps and distributed along the step edge [30]. The elastic interaction between steps thus scales as $\ln\ell$ where $\ell$ is the inter-step distance (see [30] and appendix A).

(ii) For vicinal faces with $\langle 100 \rangle$ zone axis, (case of (510) surface) the steps are parallel to the $\langle 100 \rangle$ direction (see figure 2c), the surface stress tensor reads $\frac{1}{2}\begin{pmatrix} s_{xx}+s_{yy} & s_{xx}-s_{yy} \\ s_{xx}-s_{yy} & s_{xx}+s_{yy} \end{pmatrix}$ for one terrace and $\frac{1}{2}\begin{pmatrix} s_{yy}+s_{xx} & s_{yy}-s_{xx} \\ s_{yy}-s_{xx} & s_{yy}+s_{xx} \end{pmatrix}$ for the other one (when written in the $[100]$, $[010]$ surface axis). Thus adjacent terraces have opposite surface shear stresses $\pm(s_{xx}-s_{yy})$ giving rise to a shear stress discontinuity at the step edge. This discontinuity can be described by a row of monopoles parallel to the step[1]. The monopoles of two neighboring steps are antiparallel. We show in appendix A that the elastic interaction between such monopoles also scales as $\ln\ell$. As a partial conclusion, from an elastic viewpoint, the action of the steps of such vicinal faces on the underlying crystal can be modeled by a line of elastic monopoles parallel to the step and distributed along the step edge to the usual dipolar contribution should be added.

In table 1 are reported the structural and elastic descriptions of the vicinal surfaces under study as well as the direction of the dc current (in the direction of ascending steps).

---

[1] Think about a piece of surface submitted to a shear stress that means to forces acting on each side of the elemental area and parallel to the side. When removing the half plane to form the step, only remains a net force along the step.

For completeness, notice that our description of (001) vicinal surfaces only concerns ideal surfaces. Real vicinal Si(001) surfaces misoriented towards [110] direction may exhibit a transition from single-height step to double-height step (for other zone axis, double steps have not been reported). The critical angle at which the transition occurs depends upon the sample temperature [33]. At 1150 K biatomic height steps have been found for misorientation of more than 4° [34]. Extrapolating the Fig 8 of [33], Si(113) and Si(118) surfaces should not exhibit double steps for the temperature we use.

| Vicinal surfaces | Terraces orientation | Inter-step distance a (Å) | Zone step direction | DC direction (perpendicular to the step) | Elastic description (orientation with respect to the step) |
|---|---|---|---|---|---|
| (118) | (001) | 7.7 | $[\bar{1}10]$ | $[44\bar{1}]$ | Monopoles ($\perp$) + Dipoles ($\perp$) |
| (223) | (111) | 15.6 | $[\bar{1}10]$ | $[33\bar{4}]$ | Dipoles ($\perp$) |
| (443) | (111) | 24 | $[\bar{1}10]$ | $[33\bar{8}]$ | Dipoles ($\perp$) |
| (510) | (100) | 6.75 | $[001]$ | $[\bar{1}50]$ | Monopoles (//) + Dipoles ($\perp$) |

*Table 1 :* *Description of the vicinal faces under study. The DC current is in the ascending direction to occur step bunching in the temperature range under study.*

**I.2/ Experimental procedure**

The Si single crystals of size 20x2x0.3 mm$^3$ are first chemically cleaned and then clamped between two electrodes of the sample holder in the UHV chamber. After a few flashes heating up to 1300°C during 2 minutes to clean the surface, the DC current is set to heat the sample at the chosen temperature (1100°C, 1200°C). The heating current direction is parallel to the longer side of the samples and perpendicular to the steps of the vicinal surfaces. The experiments have been performed with an ascending step current direction (regime II of [35]), for which a surface instability occurs. The current direction used for the various vicinal surfaces is reported in Table 1. Heating duration varies from 15 minutes to more than 100 hours in order to observe the whole kinetic behaviour of the faceting process. The residual pressure during heating was less than 1x10$^{-9}$ mbar. Notice that thank to the evaporation regime, a clean surface is periodically regenerated so that the surface remains clean during all the process. The samples are observed *in situ* by optical diffraction (see the experimental setup in Fig: 3a), TEM and *ex situ* by AFM and optical microscopy. The TEM apparatus is a JEOL 100C microscope modified for UHV *in-situ* experiments [36], the AFM is a Nanoscope III from Digital Instruments used in the non-contact mode. Optical diffraction experiments were performed with a laser beam (λ=0.53 nm) at an incident angle closed to the normal incidence. The scattered light is observed with a CCD camera (see fig 3a).

## II/ Experimental results

### II.1/ In-situ experiments

*Optical diffraction:*

We record the light scattered by the sample as a function of time. A few minutes are enough to obtain the diffracted pattern of Fig. 3b in the case of a Si(105) surface heated at 1200°C. The diffracted pattern reveals a periodic surface structure with a wavelength roughly around $\lambda \approx 4 \ \mu m$. We do not observe any pattern with a smaller period. Moreover, during the earlier stages of annealing (few hours), the period remains roughly fixed while the intensities of the diffracted spots change. It can thus be concluded that as soon as the sample is annealed, appears a surface undulation with a period $\lambda \approx 4 \ \mu m$, while the amplitude increases with time. For longer annealing duration (several hours) the period slowly grows towards an asymptotic stae, and the diffracted pattern is slowly blurred because of the appearance of numerous defects.

*TEM experiments:*

TEM has been used to follow the early stages of the instability. More precisely the silicon surface is illuminated in grazing incidence, so that the shadow of the edge of the sample can be observed. The amplitude of the surface corrugation is enhanced by rotating the screen in the microscope and observing the image also in grazing conditions (as described in [36]). The results, given in figure 4, show that the surface morphology exhibits a sinusoidal shape at the early beginning of the process. The time evolution of the amplitude of the corrugation is also reported in figure 4. This evolution can be perfectly fitted by an exponential law.

In other words, *in-situ* experiments point out that the early stages of annealing are characterized by the appearance of a characteristic wavelength with an exponential "explosion" of the amplitude, which are the characteristic feature of linear instability with a unique most unstable mode.

### II.2/ Ex-situ experiments

*Ex situ* experiments essentially consist in « post-mortem » examination of the samples. More precisely, the vicinal surfaces are heated in UHV, then taken out of the chamber and observed by AFM, optical microscopy and mechanical microprofilometry (Dektak 6M stylus profiler from Veeco). A set of AFM images measured from the Si(105) annealed at 1200°C during different heating duration (1h, 4.5h, 24h, 64h) is reported in Fig.5. For each picture are also reported the profiles recorded along the dotted lines. Finally, in Fig 5e 3D picture of the surface after 64 h of annealing is shown. For the shortest annealing duration (Fig. 5a), we observe some local inhomogeneities on the surface, which locally disturb the surface morphology. These points do not behave as nucleation sites since the surface exhibits also a regular wavy pattern underneath. As the heating duration increases, the one-dimensional array of bunches gets more pronounced as the size of the bunches increases. At the longest duration the bunches look more asymmetric and form microscale facets. An increase of surface disorder is also observed.
The crystallographic angles formed by the facets have been measured directly on the profiles.

### II.3/ Summary of the experimental data:

All the results are summarized in figure 6. For the Si(105) annealed at 1200°C, we have shown: the time evolution of the period (Fig. 6a), the amplitude (Fig. 6b) and the angle (Fig. 6c) of the facets. The angles are measured by AFM, mechanical microprofilometry and optical microscopy. In Fig 6d are also reported the results obtained for the various vicinals faces for two different temperatures. Notice that the initial wavelength (encircled in Fig. 6a) is roughly the same ($\approx 4\ \mu m$) whatever the initial vicinality angle while the asymptotic value (surrounded by an ellipse in Fig. 6a) depends upon the vicinal angle. Moreover, three different regimes, with peculiar characteristics are clearly observed:

(i) In the early stages, a surface instability develops exponentially with time (see also Fig. 4). The corresponding wavelength is roughly equal to $\lambda \approx 4\ \mu m$. *In-situ* optical diffraction measurements as well as TEM measurements clearly show that no smaller periodicity is observed at shorter times. This result highlights the fact that simple mechanisms based on step-pairing then double-steps pairing and so on… (e.g. zipping mechanisms.) as described in [8, 37] are not appropriate to describe the underlined mechanism. Our opinion, reinforced by *ex-situ* AFM images is that the instability proceeds by a collective motion of the steps, giving birth to a step density wave. Curiously the value of the wavelength is roughly the same whatever the initial vicinal surface (see Fig 7). In other words, at first order, this value does not depend upon the initial distance between the monoatomic steps on the initial vicinal surface. Some other authors have yet noticed that the initial wavelength roughly does not change with the vicinality angles [19-20]. In Fig 7 we report our results (stars). We can thus define a domain (the upper dotted segment in Fig. 7) in which the wavelength roughly does not depend upon the vicinality. Some authors have also studied the wavelength change versus the vicinality, so that we can report in Fig 7 two other domains where the wavelength seems to be constant whatever the initial inter-step distance. These domains are also drawn as dotted segments in Fig. 7. The three dotted segments do not merge in a single dotted line because the experiments have not been performed at the same temperature while the wavelength depends on the sample temperature [19-25]. For completeness we also report some other values "gleaned" in literature [21-25] but for which the experimental conditions (temperature, annealing duration, vicinal angle) are not well known. In any way, all these values belong to the range 1.5 $\mu m \leq \lambda \leq 5$ $\mu m$ while the vicinality angle varies by two orders of magnitude. For completeness, notice that some authors have reported some weak angle dependence [19,20,38].

(ii) At latter stages, the kinetics of faceting becomes slow, and a hill-and-valley structure forms. The bunches then start to form facets which crystallographic orientation can be easily obtained from angle measurements. It is found that the angle ($\alpha$) of one of the microscale facet remains constant while the other ($\beta$) increases with time.

(iii) After a long time, a stationary state is reached. It is formed by the $F_1$ and $F_2$ facets which crystallographic indexes are reported in table 2 for each initial vicinal face. The crystallographic nature of the facets shows that the bunches evolve towards the closest densely packed crystallographic planes surrounding the initial vicinal surface in the equilibrium crystal shape [31]. Notice that the $F_2$ facets are not flat at the atomic scale because it is easier to reach a stepped face than a flat one for which supplementary activation energy is needed for step coalescence. At the end of the first regime there is a unique wavelength but the step density still depends upon the initial vicinality. Notice that further annealing by an alternative current of the so-facetted structure restores the flatness of the nominal vicinal surface as it should be for electromigration-induced faceting.

| **Vicinal face** | **(118)** | **(223)** | **(443)** | **(510)** |
|---|---|---|---|---|
| $F_1$ (flat at the atomic scale) | (001) | (111) | (111) | (100) |
| $F_2$ (exhibit monoatomic steps) | (113) | (113) | (110) | (110) |

*Table 2:* Decomposition of the vicinal faces in $F_1$ and $F_2$ facets for the stationary state. The inter-step distance calculated in the (113) and (110) surfaces are estimated (from a projection of the inter-plane distance) to 2.88 Å and 4.46 Å respectively.

## III/ Discussion:

To sum up, all these experimental results are compatible with a quick step density wave mechanism followed by a much slower step bunching mechanism as mentioned in a previous paper [39], and as proposed in the case of non-transparent steps in Ref. [40]. During step bunching, the angle $\beta$ (defined in Fig. 6) of the microscale facet increases with time while the terraces of the initial vicinal surface remain flat ($\alpha$ is constant). The final state is a stationary state formed with the two closest facets in the equilibrium shape surrounding the initial vicinal face (see Fig. 1a) and thus is fixed by crystallography. A sketch of the mechanism of kinetic faceting (with t the time) is reported in figure 8.

In the following, we will focus on the final and initial stages of the process.

### III.1/ Analysis of the final state: towards a Marchenko-Alerhand description

Let us consider the usual faceting transition: an unstable surface (with thus negative surface stiffness) decomposes into a periodic sequence of facets with orientations $\theta_1$ and $\theta_2$ having different surface stresses [41]. The unstability originates in the decrease of the total surface energy from the planar to the facetted state. The slopes of the facets are given, but the period of the sequence is fixed by elasticity [27,29]. The surface stress discontinuities at the boundaries can be modelled by rows of monopoles perpendicular to the discontinuities [27,29]. The elastic relaxation induced by these forces diverges logarithmically [27,29] so that the elastic relaxation overcomes the energy of the domain boundaries. This results in the spontaneous formation of periodic facets with period $L$ [27,29]. More precisely, the total energy change from the flat towards the facetted state classically reads [30]:

$$\Delta E = \Delta E_{surf} + \Delta E_{bound} + \Delta W_{elast}$$

where $\Delta E_{surf}$ is the surface energy change (negative since the initial surface is unstable), $\Delta E_{bound}$ the boundary energy (positive) and $\Delta W_{elast}$ the elastic relaxation (negative).

Notice that while $\Delta E_{bound}$ and $\Delta W_{elast}$ depend upon $L$, it is not the case of $\Delta E_{surf}$ which only depends upon the crystallographic orientation of the facets. The equilibrium period fixed by the condition $\partial \Delta E/\partial L=0$, thus does not depend on $\Delta E_{surf}$ [30]. Other mechanisms can also lead to a selection of an average distance between bunches [42].

In the case under study, annealing the facetted structure without electromigration restores the nominal flat surface. In other words, in absence of electromigration, the final state is unstable ($\Delta E_{surf}$ is positive). The faceting thus is no more caused by the surface energy reduction but by a driving force due to the electrical field. Here we assume, as in Refs. [8,11,43] that the selection of the period remains based on the elastic relaxation whatever the origin of the destabilisation (thermodynamic or kinetics). It should mean that the electromigration field role is equivalent to define an effective surface energy change $\Delta E_{surf}^{Eff.}$ in the expression of $\Delta E$. Furthermore since the electric field does not depend upon $L$, $\Delta E_{surf}^{Eff.}$ does not play a role in the selection of the period.

Furthermore, in order to have a general picture –based on atomic steps- which applies at all times, we describe the final state as an elastic interaction between steps characterised by dipoles or monopoles rather than an interaction between microscale facets characterized by their own surface stress tensor.

The usual approach to calculate step-induced elastic field is (i) to describe the step in terms of localized forces distributions applied at the step edge, (ii) to model the action of these forces on the underlying crystal by point forces acting on a semi-infinite flat crystal, and (iii) to use the Green function to calculate the strain field and then the stored elastic energy [44]. The result is well known for the surface of a stress-free (resp: stressed) body (for a review see [30]) modeling the vicinal surface as a periodic array of 1D rows of elastic dipoles (resp: monopoles) perpendicular to the step edge. In our case, the description of the elastic interactions between the steps is more complex for two reasons: (i) as shown in section I the vicinal initial surfaces may be described by various configurations (alterned monopoles and/or dipoles), (ii) in the final state these rows rearrange to form a hill-and-valley structure characterized by two lengths: the step-step distance in a bunch and the distance in between two neighboring bunches. Thus the elastic description of the final state depends on the type of monoatomic steps (that means upon the initial vicinal surface) and on the characteristic lengths. However, even if electromigration is known to induce kinetics instability, the elastic energy we calculate is that one of the final facetted structure consisting in large (001) terraces separated by step bunches. This final state, reached for a maximum of the step density, is driven by energetic and not kinetics. Electromigration thus does not modify the interstep distance in a dense bunch.

To estimate the stored elastic energy modification arising from the faceting we will proceed in three steps: (i) calculation of the elastic interaction between two steps, (ii) calculation of the elastic energy of the facetted state, (iii) calculation of the elastic energy difference between the initial vicinal face and the facetted final state. Finally, we will compare our results to the usual Alerhand-Marchenko microscale approach [27,29]. We will see that the comparison will give access to the surface stress change close to a high index surface.

Notice that in the following, we will use isotropic linear elasticity. Indeed while isotropic elasticity fails to reproduce the displacement field induced by the steps, it is now well known that isotropic elasticity can be used for determining the elastic energy with a good accuracy [45].

### III.1.1/ Elastic interaction between steps:

As recalled in appendix A, the elastic interaction energy per unit length between two - parallel steps separated by a distance $\ell$ is well known (for a review see [30]). For elastic dipoles perpendicular to the step edge it scales as $\ell^{-2}$, while for elastic monopoles perpendicular to the step it scales as $\ln(\ell/a_0)$ where $a_0$ is a cut-off length of the order of a few atomic units. We show in appendix A that the elastic energy between two rows of antiparallel elastic monopoles also scales as $\ln(\ell/a_0)$ but with a different prefactor.

### III.1.2/ Elastic energy of the facetted surface:

The elastic energy in the facetted final configuration can be easily obtained by adequate summations of the elastic energy interactions between two parallel rows. For the sake of simplicity we will calculate separately the elastic energy due to the interaction of steps in a bunch (containing $N$ steps) and the elastic interaction between the bunches (see figure 9 for the geometrical definitions). The first term will be called intra-bunch energy, the second the inter-bunch energy. The analytical expressions of these energies are reported in tables 3 where for the sake of simplicity we separate the dipolar and the monopolar contributions. Thus in the following we consider the step-step interaction as described by dipole-dipole interaction or monopole-monopole interaction but never consider the dipole-monopole interaction.

The exact expressions can be expressed as a summation of the elastic energy between two steps over the considered configuration (intra or inter bunch). Approximated analytical expressions are obtained by (i) transforming the summation to an integral then by (ii) considering $N, M \gg 1$. In tables 3 are reported the expressions for elastic dipoles (Table 3a) and for elastic alterned monopoles (Table 3b).

| | **Intra-bunch** | **Interaction between two bunches (inter-bunch)** | **Interaction energy for an infinite periodic surface.** |
|---|---|---|---|
| **Exact expression** | $\dfrac{A_{dip}}{a^2}\sum_{i<j,j}\dfrac{1}{(i-j)^2}$ | $\dfrac{A_{dip}}{a^2}\sum_{i,j=1}^{N}\dfrac{1}{(M+(i-j))^2}$ | $\dfrac{A_{dip}}{a^2}\sum_{M}\sum_{i,j=1}^{N}\dfrac{1}{(M+(i-j))^2}$ |
| **Approximated expression** | $\dfrac{A_{dip}}{a^2}\left[N\dfrac{\pi^2}{6}-1-\ln N\right]$ | $-\dfrac{A_{dip}}{a^2}\ln\left[1-\left(\dfrac{N}{M}\right)^2\right]$ | $-\dfrac{A_{dip}}{a^2}\ln\left[\dfrac{\sin\left(\dfrac{\pi N}{M}\right)}{\left(\dfrac{\pi N}{M}\right)}\right]$ |

*Table 3a:* Elastic energy $W/L$ for dipoles. Moreover the expressions are given per unit step-length, thus the unity is an energy over surface area.

| | **Intra-bunch** | **Between two bunches (inter-bunch)** | **For an infinite pattern of bunches** |
|---|---|---|---|
| **Exact expression** | $\dfrac{A_{mon}}{a_0^2}\sum_{i<j,j=1}^{N}(-1)^{j-i}\ln\left((j-i)\dfrac{a}{a_0}\right)$ | $\dfrac{A_{mon}}{a_0^2}\sum_{i,j=1}^{N}(-1)^{j-i}\ln\left((M+j-i)\dfrac{a}{a_0}\right)$ | $\dfrac{A_{mon}}{4a_0^2}\sum_{M}\sum_{i,j=1}^{N}(-1)^{j-i}\ln\left((M+j-i)\dfrac{a}{a_0}\right)$ |
| **Approximated expression** | $\dfrac{A_{mon}}{4a_0^2}\left(2N\ln\left(\dfrac{\pi a_0}{2a}\right)-1\pm\ln N\right)$ | $\pm\dfrac{A_{mon}}{4a_0^2}\ln\left(1-\left(\dfrac{N}{M}\right)^2\right)$ | $\pm\dfrac{A_{mon}}{4a_0^2}\ln\left(\dfrac{\sin\left(\dfrac{\pi N}{M}\right)}{\left(\dfrac{\pi N}{M}\right)}\right)$ |

***Table 3b :*** Elastic energy $W/L$ for alterned monopoles. Notice that $A_{dip.}=+\dfrac{1-\nu^2}{\pi E}A^2$ but $A_{monop.}=\dfrac{(1+\nu)(1-2\nu)}{\pi E}F_y^2$ (see appendix A). the + and - sign arises respectively for N even and N odd. Moreover, the expressions are given per unit step-length, thus the unity is an energy over surface area.

In figure 10 are plotted the elastic interactions calculated numerically by performing the exact summations but without any monopole-dipole interaction. They are in good agreement with the approximated analytical expressions calculated by integration so that in the following we will use the approximated expressions. The main results are (i) the intra-bunch contributions depend linearly on the number of steps in a bunch (for monopoles or dipoles), (ii) for monopoles the intra-bunch elastic energy is smaller for $N$ even than for $N$ odd so that bunches prefer to be formed by an even number of steps, (iii) the inter-bunch analytical expressions are similar to the expressions given by Marchenko [27] and Alerhand [29] who modeled a facetted surface as a periodic pattern (period $L$) of 1D rows of elastic monopoles perpendicular to the edges (see bottom of figure 9), (iv) for bunches of monopoles the nature of the interaction (attraction or repulsion) depends upon the parity of $N$ (see the $\mp$ sign in table 3b). However, in the following we will only consider the stablest situation with even $N$ (see point (ii)). The point (iii) can be easily understood in the case of elastic dipoles perpendicular to the steps. Indeed, as in electrostatic, a ribbon of dipoles creates in the far field the same displacements as two rows of antiparallel monopoles located at the ribbon edges. For alterned monopoles it is quite similar since they behave as the $\ell$-apart components of a dipole. The main difference with the Marchenko-Alerhand microscale approach is that in our expressions the prefactor of the $\ln\left[1-\left(\frac{N}{M}\right)^2\right]$ term is proportional to the amplitude of the dipole or monopole prefactor while in the Marchenko-Alerhand approach it is proportional to the difference between the surface stress of the adjacent facets [27,29]. We will see in section III.2.4 that the comparison between the nanoscale and the microscale models enables us to propose an analytical expression of the surface stress angular dependence close to a high index facet.

**III.1.3/** Elastic energy change due to faceting

Let us consider the energy change due to the faceting process that means the energy change due to the transformation from a vicinal surface towards a facetted system. This energy change per unit length reads:

$$\frac{\Delta W}{L}=\Delta f(p)+\frac{\tau}{L} - \frac{\overline{A}}{L}\ln\left(\frac{L}{\pi a}\sin(\pi p)\right) \qquad (1)$$

where $\overline{A}=A_{dip}/a^2$ and $\overline{A}=A_{monop.}/4a_0^2$ for dipoles or monopoles respectively

The first term in (1) is the elastic energy change due to the step coalescence. It can be written as is a simple function $\Delta f(p)=f(p)-f(p_0)(1-p)+f(p_1)p$ where $p=N/M$ is the relative coverage of one phase with respect to the other (see figure 9) and $p_0$ and $p_1$ the slopes of the facets $F_1$ and $F_2$. The exact analytical form of $f(p)$ depends upon the monopolar or dipolar nature of the step but this is not essential. More important is the fact that $f(p)$ does not depend on the period $L=Ma$. The term $\tau$ has been introduced to take into account the boundary energy between both domains. It does not appear naturally in the simple sketch given in figure 8 but should appear when considering that because of the symmetry breaking, the steps located at the edges of the bunch cannot have the same energy as the steps inside the bunch. Finally, the last term describes the inter-bunch elastic interaction. It does not depend upon the nature of the step interaction excepted the prefactor.

When considering that the surface occupation of each domain is, at least for the final state, fixed by crystallography (since the facets in the final state correspond to cusps of the gamma-plot [31]), the energy change per unit length is a simple function of $L$, so that its minimum value is reached for $\left.\frac{\partial \Delta W/L}{\partial L}\right|_p = 0$. The equilibrium period thus reads:

$$\lambda = \frac{\pi a}{\sin(\pi p)} \exp\left(\frac{\tau}{A}\right) \qquad (2)$$

This expression can be compared to that obtained by Marchenko [27] and Alerhand [29]. They considered the final state as formed by microscale facets characterized by their own surface stress whose components perpendicular to the facet edges are drawn in Fig. 8 at $t=t_\infty$:

$$\lambda = \frac{\pi a_0 e}{\sin(\pi p)} \exp\left(\frac{\pi E \tau}{2(1-\nu^2)|\vec{s}_1 - \vec{s}_2|^2}\right) \qquad (3)$$

The fit of the experimental results give the ratio $\tau/A$ for the vicinal surfaces under study (see table 4 where the value of $a$ have been estimated from table 2)

|       | $\lambda$ (Å)   | $p = \tan\theta_1/(\tan\theta_1 + \tan\theta_2)$ | $\tau/A$ |
|-------|-----------------|--------------------------------------------------|----------|
| (118) | 6.6 10³         | 0.39                                             | 6.50     |
| (223) | 12 10³          | 0.38                                             | 7.10     |
| (443) | 13 10³          | 0.19                                             | 6.25     |
| (510) | 8 10³           | 0.23                                             | 5.93     |

**_Table 4_** : *Experimental values of the period, the relative occupation then the so-deduced ratio $\tau/A$ values for T=1150°C*

Thus, within the experimental error bars, we find at T=1150 °C, $\langle \tau/\overline{A} \rangle \approx 6.5 \pm 0.5$ whatever the initial surface. More precisely, for the (223) and (443) surfaces (for which the steps only bear elastic dipoles) one obtains, when using $A_{dip} \approx 10^{-30}$ $J.m$ (see section III.1), $\tau \approx 8.6 \ 10^{-11}$ $Jm^{-1}$ for the (223) and $\beta \approx 3.1 \ 10^{-11}$ $Jm^{-1}$ for the (443) surface. For the (118) and (510) surfaces the steps bear elastic dipoles and monopoles (see table 1) so that we cannot simply extract $\tau$ from table 4. Indeed as reported at the beginning of the section III.1.2 the dipolar and monopolar contributions do not simply add so $\overline{A}$ is an unknown composition of $A_{dip}$ and $A_{mon}$. However if the amplitude of the monopoles can be neglected with respect to the amplitude of the dipole we get $\tau \approx 7.9 \ 10^{-11}$ $Jm^{-1}$ for the (118) surface and $\tau \approx 3.0 \ 10^{-11}$ $Jm^{-1}$ for the (510) surface. On the contrary if we only consider the monopolar contribution, with $A_{monop}/a_0^2 \approx 3.10^{-12}$ $J.m^{-1}$ (see section III.1), there is $\tau \approx 2 \ 10^{-11}$ $Jm^{-1}$ whatever the vicinal surface under consideration. Notice that in both cases (monopoles or dipoles) (i) the order of magnitude of $\tau$ is comparable to the step energy reported for the Si(111) surface ($3.10^{-11} Jm^{-1}$) [46] and that (ii) when considering only the dipolar contribution we obtain two set of values, one around $\tau \approx 8 \ 10^{-11}$ $Jm^{-1}$ when the facet edges separate a (001) or (111) from a (113) facet, the other around $\tau \approx 3.0 \ 10^{-11}$ $Jm^{-1}$ when the facet edges separate a (001) or (111) from a (110) facet (see table 2).

**III.1.4/ Link between the nanoscale and the microscale model: the surface stress angular dependence**

The nanoscale and microscale models are equivalent if the cut-off length $a_0$ of the microscale model depends upon the initial inter-step distance $a$ (more precisely $a_0 e = a$) and if from (2) and (3) we can write the equality

$$\overline{A} = 2\frac{1-\nu^2}{\pi E}(\vec{s}_1 - \vec{s}_2)^2 \qquad (4)$$

with again $\overline{A} = A_{dip}/a^2$ and $\overline{A} = A_{monop}/4$ for dipoles and monopoles respectively.

For dipoles, introducing the step height $h$ (so that $a = h/\tan\theta$ where $\theta$ is the angle of the vicinal facet) and using $A_{dip} = 2\frac{1-\nu^2}{\pi E}A^2$ (compare Eq. A2 to A3 in the appendix), equation (4) reads:

$$\frac{A^2}{h^2}\tan^2\theta = s_1^2 + s_2^2 - 2s_1 s_2 \cos\theta \qquad (5)$$

However, in appendix A we show that (see Eq. A2):

$$A^2 = A_1^2 + (hs_1)^2 \qquad (6)$$

where $s_1 = s_{xx}$ is the surface stress component perpendicular to the step.

For weak values of $\theta$ one obtains from comparison of the two previous relations:

$$s_2 = s_1 - \frac{|\theta|}{h}A_1 \qquad (7)$$

This expression is analogous to the one found by Salanon for stressed solids [47] where the steps are described by the sum of rows of dipoles and monopoles (both perpendicular to the step) and the surface stress expression is developed up to second order in $\theta$.

Equation (7) means that since the presence of steps leads to surface stress relaxation, the surface stress is maximum for a low index surface and thus decreases with $|\theta|$. On the contrary, the energy cost to create surface steps implies that the surface energy increases with $|\theta|$. In other words, local minima (cusps) of the surface energy plot (gamma-plot) correspond to local maxima (anticusps) of the surface stress plots [30,39,47].

Beyond this approach it is also possible to use our experimental results to obtain absolute values of surface stress. Indeed, using the microscale model of Marchenko-Alerhand [27,29], the measurement of the final period gives the difference $(s_1 - s_2)$ between the surface stress components (normal to the step) of the facets $F_1$ and $F_2$. Using a set of vicinal surfaces (labeled $k$) chosen to form a closed cycle on the stereographic projection, we measure $\lambda^k(s_1^k, s_2^k)$ and thus obtain a set of values $s_1^k - s_2^k$ corresponding to the surface stress differences between the facets $F_1$ and $F_2$ that appear on the vicinal faces $k$. Since working on a closed cycle, the measurement of the periods $\lambda^k$ is enough to obtain the absolute values $s_1^k$ and $s_2^k$. The method has been extended to all the intermediate facetted stages obtained after a time $t$ smaller than the duration needed to reach the final state. In this case it is necessary to measure the period $\lambda^k(s_1^k, s_2^k)$ as well as the angles $\alpha_t^k, \beta_t^k$ formed by the facets obtained at $t$ and then to solve the systems of equations $\lambda_t^k(s_{1,t}^k, s_{2,t}^k, \alpha_t^k, \beta_t^k)$ to obtain the values $s_{i,t}^k$ of the facets $\alpha$ and $\beta$ appearing at time $t$ and characterized by the angles $\alpha_t^k$ and $\beta_t^k$. Many numerical solutions exist but only one set of $s_i^k$ values verifies the fact that all the faces that belong to the equilibrium shape exhibit a maximum of surface stress. This procedure has been used to obtain, for the first time, the complete surface-stress plot of Si [39].

**III.2/ Analysis of the initial step density waves: towards a unique wavelength $\lambda \approx 4 \ \mu m$**

In this section, we discuss the origin of the robustness of the wavelength of the initial step density waves with respect to the vicinality of the original surface. Most of the previous models concerning the step bunching instability are based on the Stoyanov approach of the step bunching instability induced by the electromigration [16]. More precisely, different regimes have been studied to calculate the most unstable mode in the linear regime. For slow attachment kinetics the main period of the instability depends upon the transparency parameters at the steps and reads $\lambda = 2\pi a_0^{-1} (6A\xi)^{1/2} \left( \dfrac{a_0}{\ell} \right)$ for opaque or moderately transparent steps [18] and $\lambda = 2\pi a_0^{-1} (6A\xi)^{1/2} \left( \dfrac{a_0}{Q\ell^{1/2}} \right)$ for very transparent steps [40]. In both expressions $a_0$ is an atomic distance unit, $A$ is an elastic quantity describing the dipolar forces at the steps, $\xi$ the reduced electromigration force and $Q$ a characteristic length varying from a tenth of an atomic distance up to some atomic distances [40]. Both expressions can be put in the generic form of a characteristic lengthscale $\lambda = 2\pi a_0^{-1} (6A\xi)^{1/2}$ times a "geometric factor", which is a dimensionless combination of atomic scales. Indeed, the interstep distance $\ell$ is of the order of some atomic distances in the experiments presented above. We here show that this generic form can be derived within the frame of a continuous model, which does not refer to microscale details.

For this purpose, we consider a model in which the initial surface is rough since the vicinal surfaces under study have high slopes. We write a continuum model based on macroscopic quantities having smooth orientation dependence. We use a 1D model, along the variable $x$ and we neglect sublimation or growth.

From the mass conservation equation:

$$\frac{\partial h}{\partial t} = - \frac{\partial j}{\partial x} \qquad (8)$$

with $h$ is the local height and $j$ is the surface flux.

We then consider the diffusion process driven by the variations of the chemical potential $\mu$ and the electromigration force:

$$j=\frac{M}{kT}c\left(f-\frac{\partial \mu}{\partial x}\right) \quad (9)$$

where $M$, $f$ and $c$ are respectively the orientation-dependent mobility, migration force, and concentration of adatoms at the surface.

We write the free energy of the surface as:

$$F=\int \varphi(p)dx=\int\left(\gamma_0+\gamma_1\left|\frac{\partial h}{\partial x}\right|+\gamma_3\left|\frac{\partial h}{\partial x}\right|\right)dx \quad (10)$$

where $\gamma_0$ and $\gamma_1$ are constants, and $\gamma_3$ is a function of the local slope $p=\partial h/\partial x$. More precisely, for usual vicinal surfaces described as a 1D array of elastic dipoles $\gamma_0$ is the terrace energy, $\gamma_1$ is the step energy and $\gamma_3=\beta_3(\partial h/\partial x)^2$ depends upon the step-step interaction energy $\beta_3$ and is proportional to the square of the local slope (see appendix A) so that $\varphi(p)$ reads: $\varphi(p)=\gamma_0+\gamma_1|p|+\beta_3|p|^3$. In the following, we consider $\gamma_3$ as a simple unknown function of the local slope $p=\partial h/\partial x$ to take into account for different types of vicinal surfaces.

The chemical potential is defined as

$$\mu=\frac{\delta F}{\delta h}\left(\frac{\delta N}{\delta h}\right)^{-1} \quad (11)$$

where $N=\int h/a_0^2 dx$ is the number of atoms of the solid.

A variational calculation then leads to the usual Herring expression [48] of the surface energy variation:

$$\delta F=-\int\frac{\partial}{\partial x}\frac{\partial \varphi}{\partial p}\delta h \quad (12)$$

so that for a positive slope one obtains:

$$\mu=-a_0^2\tilde{\gamma}_3\left(\frac{\partial h}{\partial x}\right)\left(\frac{\partial^2 h}{\partial x^2}\right) \quad (13)$$

with $\tilde{\gamma}_3=2\frac{\partial \gamma_3}{\partial p}+\frac{\partial^2 \gamma_3}{\partial p^2}p$

Using Eq. (13) into Eqs. (8) and (9), we obtain the time evolution equation of the surface height as:

$$\frac{\partial h}{\partial t}=-\frac{\partial}{\partial x}\left[\frac{Mc}{kT}\left(f+\frac{\partial}{\partial x}\left(\tilde{\gamma}_3 a_0^2\frac{\partial^2 h}{\partial x^2}\right)\right)\right] \quad (14)$$

For small height perturbations around the mean orientation of the vicinal surface of average slope $\bar{p}$, we have $h=\bar{p}x+\delta h$ which leads to:

$$\partial_t h = -\partial_p \left[\frac{Mcf}{kT}\right]_{p=\bar{p}} \partial_{xx}\delta h - a_0^2 \left[\tilde{\gamma}_3 \frac{Mc}{kT}\right]_{p=\bar{p}} \partial_{xxxx}\delta h \qquad (15)$$

where the partial derivatives are noted $\partial h/\partial i = \partial_i h$.

Considering in eq. (15) a wavelike perturbation of the height $\delta h = \exp(i\omega t + ikx)$ leads to the following equation:

$$i\omega = \partial_p \left[\frac{Mcf}{kT}\right]_{p=\bar{p}} k^2 - a_0^2 \left[\frac{\tilde{\gamma}_3 Mc}{kT}\right]_{p=\bar{p}} k^4 \qquad (16)$$

A criterion for the bunching instability to occur is that the prefactor of the term in $k^2$ should be positive. The maximum growth rate is reached for:

$$\lambda = 2\pi \sqrt{\left(\frac{2\tilde{\gamma}_3 a_0^2 [Mc]_{p=\bar{p}}}{\partial_p [Mcf]_{p=\bar{p}}}\right)} \qquad (17)$$

Let us now separate the amplitude from the angle dependence of $M$, $c$ and $f$. For this purpose we define:

$$M = M_0 g_M(p), \quad c = c_0 g_c(p) \quad \text{and} \quad f = f_0 g_f(p) \qquad (18)$$

where $g_i(p)$ are dimensionless functions of the order of one.

The wavelength then reads:

$$\lambda = 2\pi \sqrt{\frac{2\tilde{\gamma}_3 a_0^2}{f_0}} \sqrt{\frac{g_c g_M}{\partial_p (g_M g_c g_f)_{p=\bar{p}}}} \qquad (19)$$

where we omit, for the sake of simplicity, the $p$ dependence by writing $g_i(\bar{p}) = g_i$.

An inspection of Eq. (19) shows that the wavelength does not depend on the amplitude $M_0$ of the mobility neither on the amplitude of the mobile concentration $c_0$.

It is important to note that since the vicinal surfaces at high slopes are far from singular facets, the orientation dependences $g_i(p)$ do not exhibit any singularities, so that the last term of the previous relation must have a weak slope dependence.

Let us discuss more precisely the different terms of Eq. (19). In absence of growth or sublimation, the mobile adatom concentration should be at equilibrium $c_0 = c_{eq}$ so that $g_c(p) = 1$. In this case there are two possible expressions of the wavelength according to the $p$-dependence of $\tilde{\gamma}_3$ that means according to the monopolar or dipolar nature of the steps. The results are summarized in table 5.

|  | $\widetilde{\gamma}_3$ | $\lambda$ |
|---|---|---|
| **Dipoles** | $A_{dip.}\pi^2 p/a_0^2$ | $2\pi^2\sqrt{\dfrac{2A_{dip.}}{f_0}}\left[\dfrac{\bar{p}g_M(\bar{p})}{\partial_p(g_M g_f)\big|_{p=\bar{p}}}\right]^{1/2}$ |
| **Monopoles** | $A_{monop.}/pa_0^2$ | $2\pi\sqrt{\dfrac{2A_{monop.}}{f_0}}\left[\dfrac{g_M(\bar{p})}{\bar{p}\,\partial_p(g_M g_f)\big|_{p=\bar{p}}}\right]^{1/2}$ |

*Table 5: Expressions $\widetilde{\gamma}_3$ and $\lambda$ obtained for dipoles and monopoles. The interaction energies used for the calculations are given by equations A.4 in the appendix.*

Let us calculate the order of magnitude of the wavelength $\lambda$.

- For the dipolar case the value of the dipolar moment of Si is known to be roughly $A_{dip.}\pi^2/6 \approx 10^{-30}$ $J.m$ [49,50]. Using the electromigration force expression $f_0 = zeE_m$ where $z$ is the effective charge ($0.01 < z < 0.1$) [51,52], $e$ is the electronic charge and $E_m = 400 Vm^{-1}$, there is $8\,\mu < 2\pi^2\sqrt{\dfrac{2A_{dip.}}{f_0}} < 20\,\mu$. This result is slightly larger than our experimental value $\lambda \approx 4$ $\mu m$ so that there should be $\left[\dfrac{\bar{p}g_M(\bar{p})}{\partial_p(g_M g_f)\big|_{p=\bar{p}}}\right]^{1/2} < 1$.

- For the monopolar case $A_{monop}$ can be roughly estimated from the surface stress of the Si(001) surface. Since $s_1 + s_2 \approx 1$ $Nm^{-1}$ [29,53], and using formula of appendix A2 one obtains $A_{monop}/a_0^2 \approx 3.10^{-12}$ $J.m^{-1}$ so that $3\mu < 2\pi\sqrt{\dfrac{2A_{monop.}}{f_0}} < 11\mu$.

The fact that the wavelengths are comparable for monopoles and dipoles can be easily understood, since in elasticity the only dimensional constant is the Young modulus $E$, and the only specific length-scale is the atomic distance $a_0$. As $A_{dip.}$ and $A_{monop}$ scale as $Ea_0^4$ [54], it is thus possible, from (12), to write for monopoles and for dipoles:

$$\lambda = 2\pi\sqrt{2}\,a_0^2\left(\dfrac{E}{f_0}\right)^{1/2}\left[\dfrac{\widetilde{g}_3 g_c g_M\big|_{p=\bar{p}}}{\partial_p(g_M g_c g_f)\big|_{p=\bar{p}}}\right]^{1/2} \qquad (13)$$

where we have defined $\widetilde{\gamma}_3(p)=a_0^2 E \widetilde{g}_3(p)$. Therfore, the order of magnitude of the wavelength has to be the same for dipoles or monopoles. Nevertheless, here because of rough approximations the order of magnitude of the $A_{dip.}$ and $A_{monop}$ values appear to be larger than the experimental ones. Our experimental results are consistent with the fact that the bracket in (13) must be a very weak function of the slope, at least for the vicinal faces under study characterized by high vicinality angles.

## IV/ Conclusion:

The main characteristics of the faceting mechanism, in the transparency regime, of vicinal surfaces characterised by a high density of steps are the following:

(i) In the early stages, the instability takes the form of a step density wave, with a fixed wavelength, and amplitude that increases exponentially with time. The corresponding wavelength is roughly equal to $\lambda \approx 4 \ \mu m$. Considering a continuum model based on macroscopic quantities having a weak orientation dependence, we have shown that the order of magnitude of the wavelength does not depend upon the details of the surface at the atomic level, such as: step transparency and kinetic properties, elastic description of the initial vicinal surfaces (dipoles or monopoles), or the vicinality angle (at least to leading order).

(ii) At latter stages, the kinetics of faceting becomes slow, and a hill-and-valley structure form by a process in which the terrace orientation is conserved but the facet orientation increases with time. We have not studied in detail the kinetics of the mechanism, which will be reported in a fore-coming paper.

(iii) Asymptotically, a stationary state is reached. The stationary facets are the closest densely packed crystallographic planes surrounding the initial vicinal surface in the equilibrium crystal shape. Because of the activation energy needed for step coalescence the facets ($F_2$) are not flat at the atomic scale, while the terraces ($F_1$) remain flat at the atomic scale. For both situations (dipoles or monopoles), the final state was described in terms of energetic competitions between elastic relaxation and the cost need to create the facets edges, as described by Marchenko [27] then Alerhand [29] by using directly a microscale model. The comparison between the analytical expressions issued from the two approaches: step models and the microscale approaches gives access to the angular dependence of the surface stress. This can be used to study the surface stress anisotropy as first reported in [39].

Last but no least, our results show that it is possible to tune the period of the faceting in the micrometric range. The goal now is to be able to tune the faceting at the nanoscale. It could be possible by using growth instability [55] or externally applied stress [56].

**Acknowledgements:**

This work has been supported by the ANR PNANO grant: Nano-morphogénèse. The authors would like to thank A. Kalifa for his help to use microprofilometry, F. Pailherey for his technical help and B.Croset, A.Saùl and R. Kern for helpful discussions.

# Appendix:
## Elastic interaction between steps

The elastic energy stored in an elastic body is simply half the work done by the surface force distribution $\vec{P}(\vec{x})$ (characterized by its components $P_\alpha(\vec{x})$) against the surface displacement. It can be written (for a review see [30]):

$$W = \frac{1}{2} \sum_{\alpha,\beta} \iint P_\alpha(\vec{x}) D_{\alpha\beta}(\vec{x},\vec{x}') P_\beta(\vec{x}') d^3x\, d^3x' \quad \text{(A1)}$$

where $D_{\alpha\beta}(\vec{x},\vec{x}')$ (with $\alpha,\beta = x,y$) is the Green tensor that means the displacement field $\vec{u}(\vec{x})$ associated to a point force of amplitude unity located at $\vec{x}'$ [44].

For two (parallel or antiparallel) monopoles located in $(x_1,y_1,0)$ and $(x_2,y_2,0)$ the force distribution reads [30] $P_\alpha(\vec{x}) = F_\alpha [\delta(x-x_1)\delta(x-y_1) \pm \delta(x-x_2)\delta(x-y_2)]\delta(z)$ with the sign + for parallel monopoles and − for antiparallel monopoles [2] where $F_\alpha$ has the dimension of a force ($\alpha = 1,2,3$).

For two (parallel or antiparallel) dipoles perpendicular to the $y$-direction (parallel to the step), the force distribution reads [30]: $P_\alpha(\vec{x}) = A_\alpha \left[ \left.\frac{\partial \delta(x)}{\partial x}\right|_{x-x_1} \delta(x-y_1) \pm \left.\frac{\partial \delta(x)}{\partial x}\right|_{x-x_2} \delta(x-y_2) \right]\delta(z)$

where $A_\alpha$ has the dimensions of a mechanical torque. Here again, the sign + is for parallel dipoles and − for antiparallel dipoles. For a step dividing the surface in two equivalent terraces, the surface stresses $s_{11}$ of the two neighboring terraces exert a mechanical torque per unit length of moment $s_{11}h\hat{y}$ ($h$ being the step height) which has to be equilibrated by the torque of the force dipolar distribution so that [57,58,30] $A_3 = s_{11}h$. On the contrary, the $A_1$ component can only be calculated by means of inter-atomic potentials (see for example [59]).

Let us now consider two $\ell$-apart steps parallel to the $\vec{y}$ direction bearing identical dipoles parallel to the $\vec{x}$ direction or antiparallel monopoles in the $\vec{y}$ direction
Using the properties of the Dirac "function" there is for the monopoles

$$W_{monop.} = F_y^2 \int_{-\infty}^{\infty}\int_{-\infty}^{\infty} [D_{yy}(0,y-y') - D_{yy}(\ell,y-y')] dy\, dy'$$

---

[2] $\delta(x)$ is the Dirac "function"

and for the dipoles

$$W_{dip.}=A^2\int_{-\infty}^{\infty}\int_{-\infty}^{\infty}\left[\left.\frac{\partial D_{x,x}}{\partial x}\right|_{0,y-y'}+\left.\frac{\partial D_{xx}}{\partial x}\right|_{\ell,y-y'}\right]dydy'$$

where $D_{ii}=\frac{1-\nu^2}{\pi E}\left[\frac{1}{r}-\frac{\nu}{1-\nu}\frac{(i-i')^2}{r^3}\right]$ with $r=\sqrt{(x-x')^2+(y-y')^2}$ for $i=x,y$ and now $A^2=A_1^2+A_3^2$

Performing thus the integral there is, when defining the density of elastic energy of interaction per unit length of the step $w=\lim_{L\to\infty}\frac{W}{L}$:

$$w_{dip}=+2\frac{1-\nu^2}{\pi E}A^2\frac{1}{\ell^2}\text{ with }A^2=A_1^2+(s_{11}h)^2 \qquad w_{monop.}=-2\frac{(1+\nu)(1-2\nu)}{\pi E}F_y^2\ln\left(\frac{\ell}{a_0}\right) \qquad \textbf{(A2)}$$

where the quantity $a_0$ is an atomic unit introduced as a cutoff in order to avoid local divergences in the calculation of the integrals. In both cases the rows repulse each other. Notice that $w_{monop}$ diverges while $w_{dip}$ converges.

Let us note that when performing the same summation for parallel monopoles in the $\vec{x}$ direction one recovers the well-known result: $w_{monop.}=+2\frac{1-\nu^2}{\pi E}F_x^2\ln\left(\frac{\ell}{a_0}\right)$.

In the following, equations (A2) will be written:

$$w_{dip.}=\frac{A_{dip}}{\ell^2} \qquad \text{and} \qquad w_{monop.}=-\frac{A_{monop}}{a_0^2}\ln\left(\frac{\ell}{a_0}\right) \qquad \textbf{(A3)}$$

Notice that with these notations, $A_{dip}$ and $A_{monop}$ have the same units: energy times length.

The previous results (A3) can now easily be extended to the case of an infinite array of parallel rows. For this purpose it is enough to use the superposition principle and thus to do the corresponding summations. For vicinal surfaces formed by a periodic pattern of parallel rows, the results simply reads:

$$w_{dip.}^{Vic.}=\frac{\pi^2}{6}\frac{A_{dip}}{\ell^2} \qquad \text{and} \qquad w_{monop.}^{Vic}=-\frac{A_{monop}}{2a_0^2}\ln\left(\frac{2\ell}{\pi a_0}\right) \qquad \textbf{(A4)}$$

For facetted surfaces formed by step bunches separated by flat terraces the summations are less easy to perform. They are given in the tables 3 where are also given the approximated expressions obtained by substituting integrals to sums. The exact and approximated expressions are compared in figure 10.

# Figure captions

**Figure 1 :** **a)** Stereographic representation of the vicinal surfaces under study. The arrows represent the normal to the vicinal surfaces. Notice that the (001), (113), (111), (110) and (100) surfaces belong to the Silicon equilibrium shape [31,39]. **b)** projection along the [-110] direction of some of the studied vicinal surfaces of a cfc material. Notice that the (111) and (001) surfaces are flat at the atomic level, that the (113) and (101) surfaces are flat at the second neighbour (atomic stepped surfaces). For the sake of simplicity we only consider in figure 2b the simple case of a cfc crystal and not the true diamond structure of the Silicon. It is enough for our purpose.

**Figure 2:** Schematic representation of the various kinds of vicinal surfaces under study. According to the terrace structure, the steps can be described as rows of elastic dipoles or by the sum of elastic monopoles and elastic dipoles. Furthermore for the vicinal of the (001) surface one step is rougher than the other [32] as drawn in Fig. 2b . The couples of dots represent the dimers and the lines on the terraces the rows of dimmers.

**Figure 3: a)** Sketch of the optical diffraction equipment, **b)** Optical diffraction pattern recorded for Si(105) after 1 hour at 1200°C. Label 0, +1, +2 and −1 correspond to the different orders of diffraction.

**Figure 4:** TEM observation of the first stages of the roughening of the (105) Si surface. The images correspond to 5 min, 1h then 2h of annealing (Notice the two different perpendicular scales due to the grazing incidence). In the bottom right part of the figure is also reported the time evolution of the amplitude of the oscillation (T =1250 C)

**Figure 5:** AFM images of the (105)Si surface evolution versus time. Figures a,b,c,d,and e respectively correspond to 1h, 4.5h, 24 h and 64 h of annealing at 1250 C The corresponding profiles (obtained along the dotted lines) are reported just below. In (e) is reported the 3D picture obtained after 64 h of annealing.

**Figure 6:** Summary of the experimental results obtained at 1250 °C. For the (105)Si surface are reported the time evolution of the period (a), of the amplitude (b) then of the angles formed by the facet (c). In fig.7d are synthesised the results obtained a set of vicinal surfaces at two different temperatures. Sections III.1 and III.2 of the discussion will be devoted to the initial and final part of the curves surrounded in (a), (b) and (c).

**Figure 7:** In thus figure are reported the wavelengths that appear at the very beginning of the process. More precisely we report our results (stars) as well as the results obtained by other authors in other contexts. The dotted lines correspond to domains in which no wavelength change (see text for more details). White circle [21], black down-triangles [25], white squares [19-20], diamonds [60], white up-triangle [23], white down-triangles [22], black cross [24].

**Figure 8:** Sketch of the faceting mechanisms. At t=0 is the initial vicinal surface. At t=$t_1$, a step wave density forms by collective motion of the steps. At t=$t_2$ the step bunching mechanism starts so that the initial terrace ($F_a$) grows at constant angle while a facet ($F_b$) forms with the angle $\beta(t_2)$. At the end of the process (t=∞) there is a stationary state formed by the flat facet $F_1$ and the stepped facet $F_2$ characterised by their own surface stress tensors. The surface stress component perpendicular to the edge are $\vec{s}_1$ and $\vec{s}_2$.

**Figure 9:** Elastic model used for the calculations. The period L=Ma consists in a flat terrace and a step bunch (inter-step distance a) formed by (N-1) steps. In the bottom part of the figure are reported the corresponding nanoscale and Marchenko Alerhand models. In the nanoscale model, steps (in the bunch) are modelled by rows of point forces (in the figure are only sketched the elastic dipoles perpendicular to the steps, at which could be added elastic monopoles parallel to the steps according to the description of the vicinal faces under consideration as shown in figure 2). In the Marchenko-Alerhand model, the bunch itself is considered as a microscale facet modelled by rows of elastic monopoles located (and perpendicular) to the facet edges.

**Figure 10:** Graph of the elastic energies reported in tables 3: (a) intrabunch term calculated for dipoles, (b) interbunch term calculated for dipoles, (c) intrabunch term calculated for monopoles (the upper curve is for N odd, the lower curve for N even), (d) interbunch term calculated for monopoles with even N.

# Figures

**Fig. 1a**

**(a)**

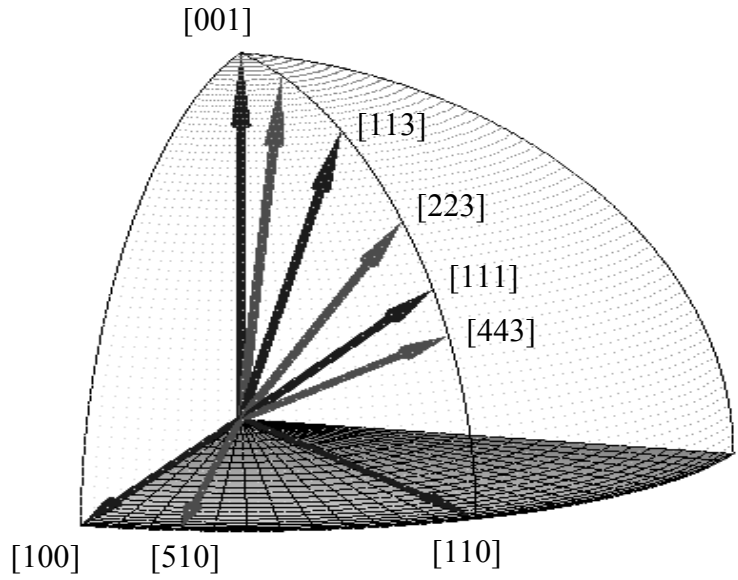

**(b)**

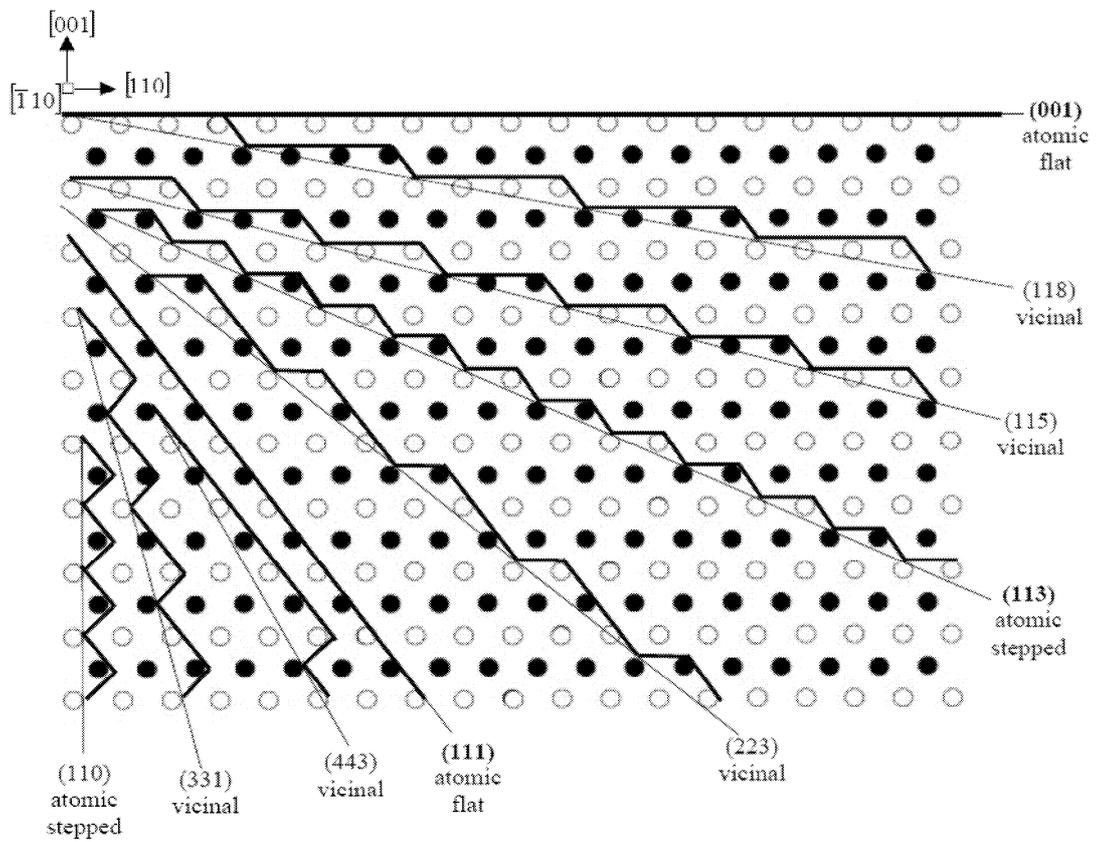

**Figure 2**

| | |
|---|---|
| (a) 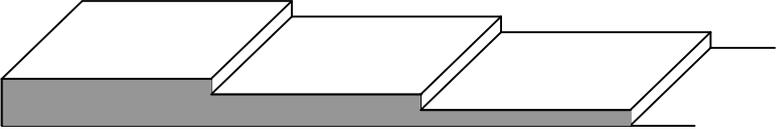 | **Vicinal (111)**<br>Identical terraces<br>Identical steps<br><br>Dipôles |
| (b) 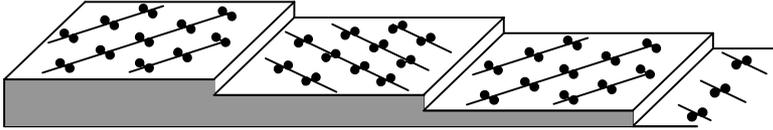 | **Vicinal (001) in the [100] direction**<br>Different terraces<br>Different steps<br><br>Monopoles |
| (c) 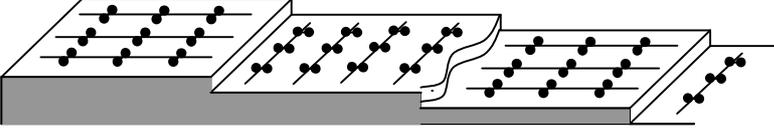 | **Vicinal (001) in the [110] direction**<br>Different terraces<br>Identical steps<br><br>Monopoles + dipoles |

Figure 3 :

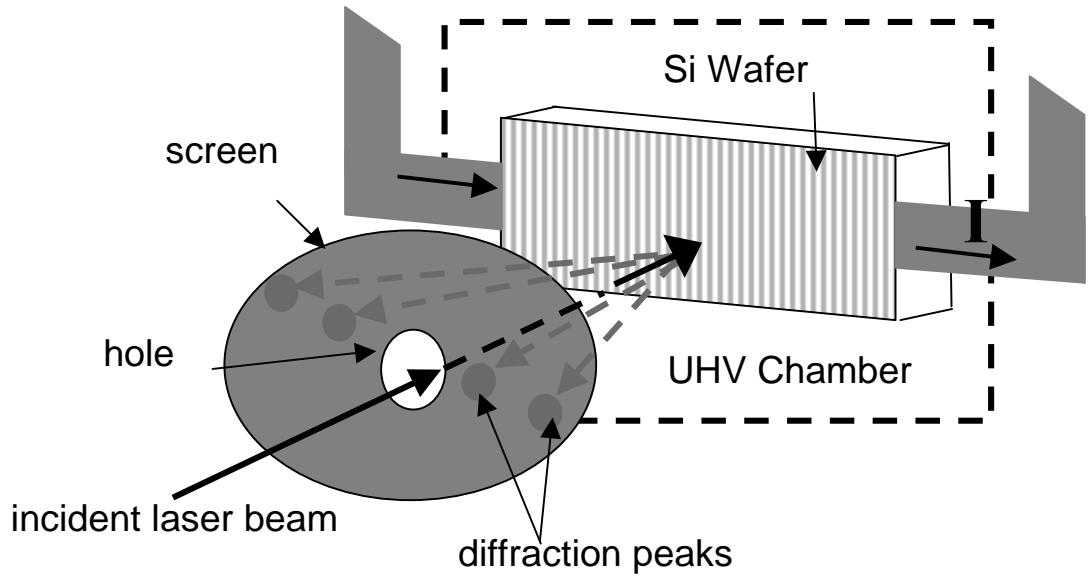

(b)

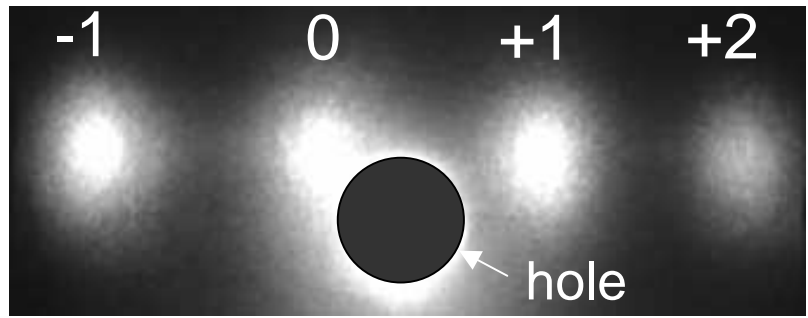

Figure 4 :

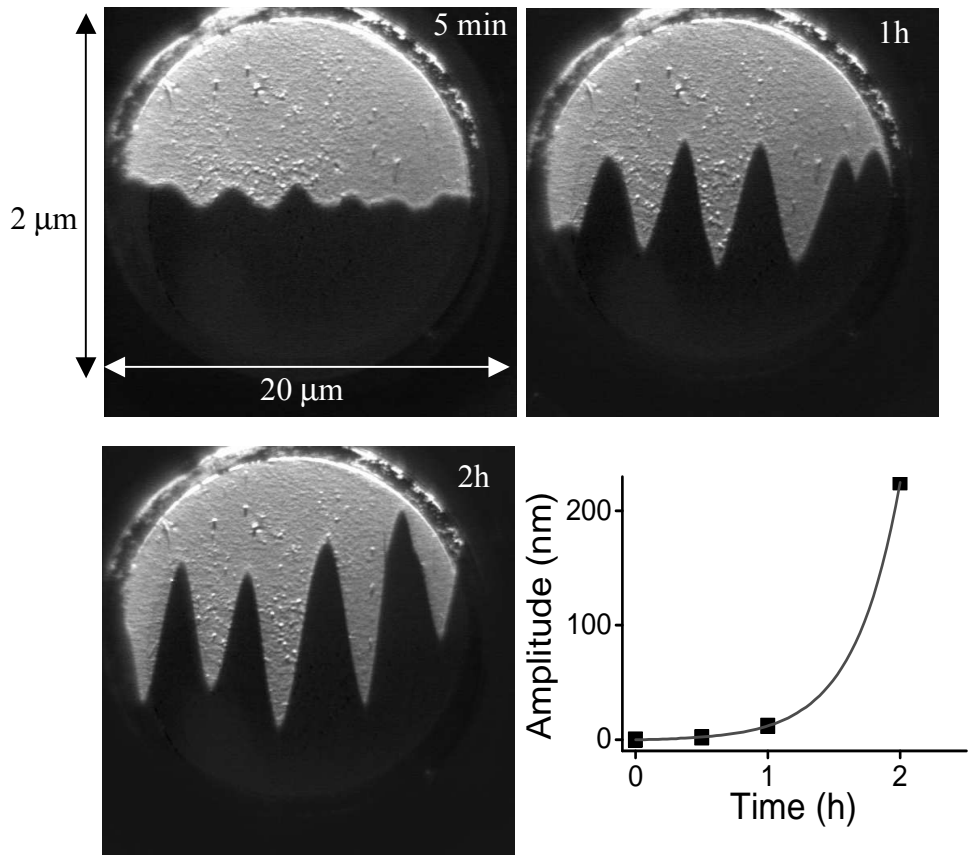

Fig 5 :

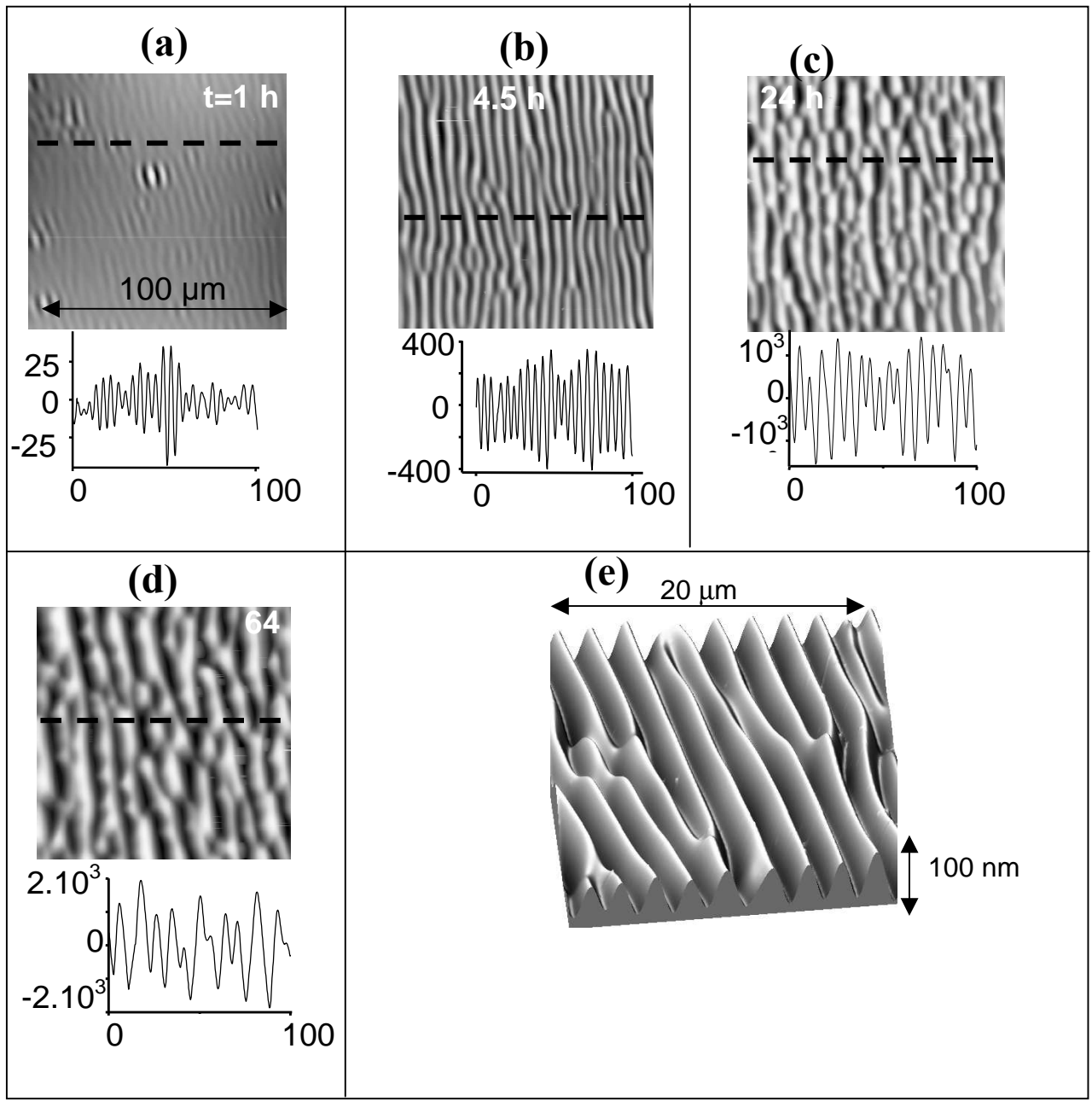

Fig 6 :

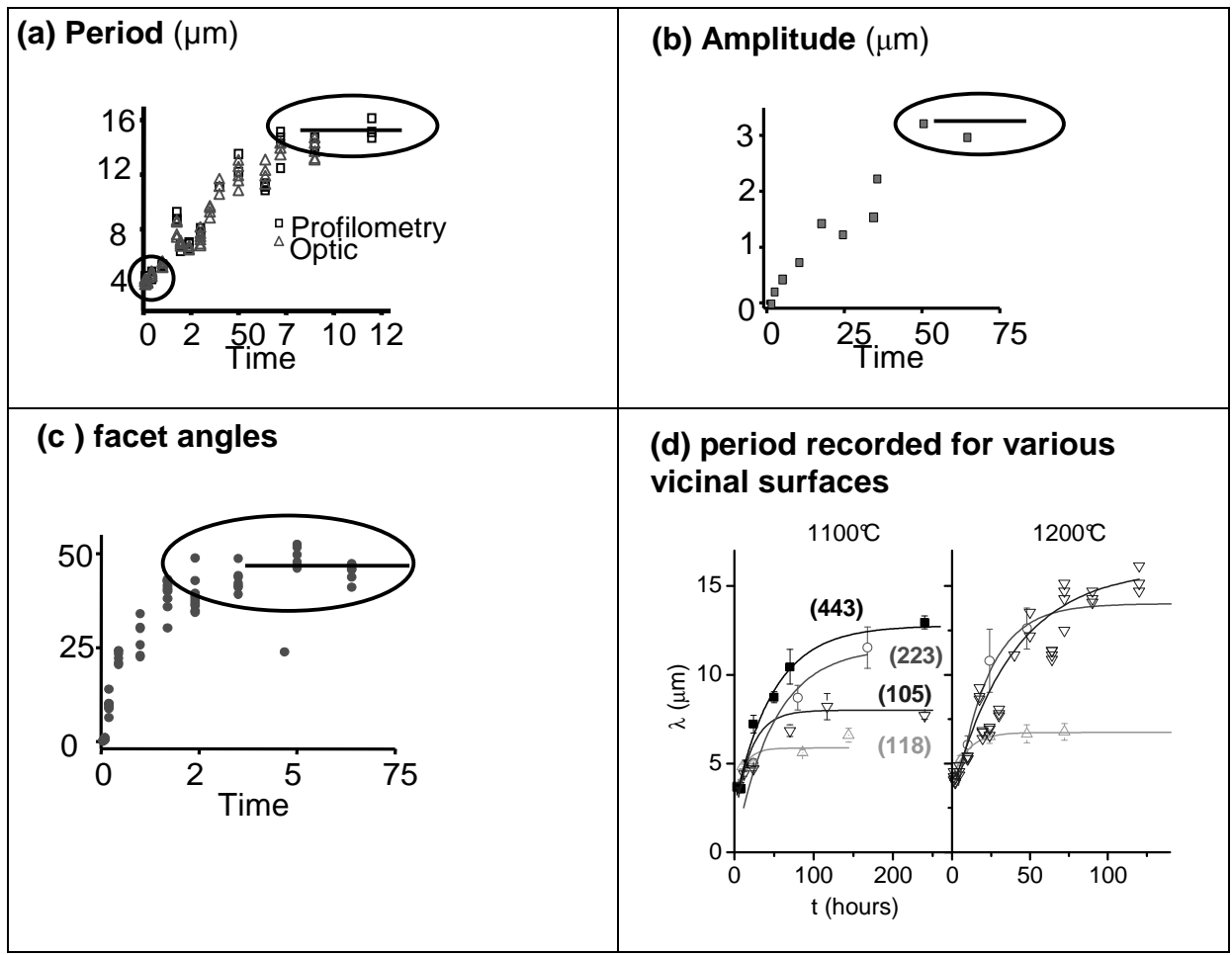

Fig 7:

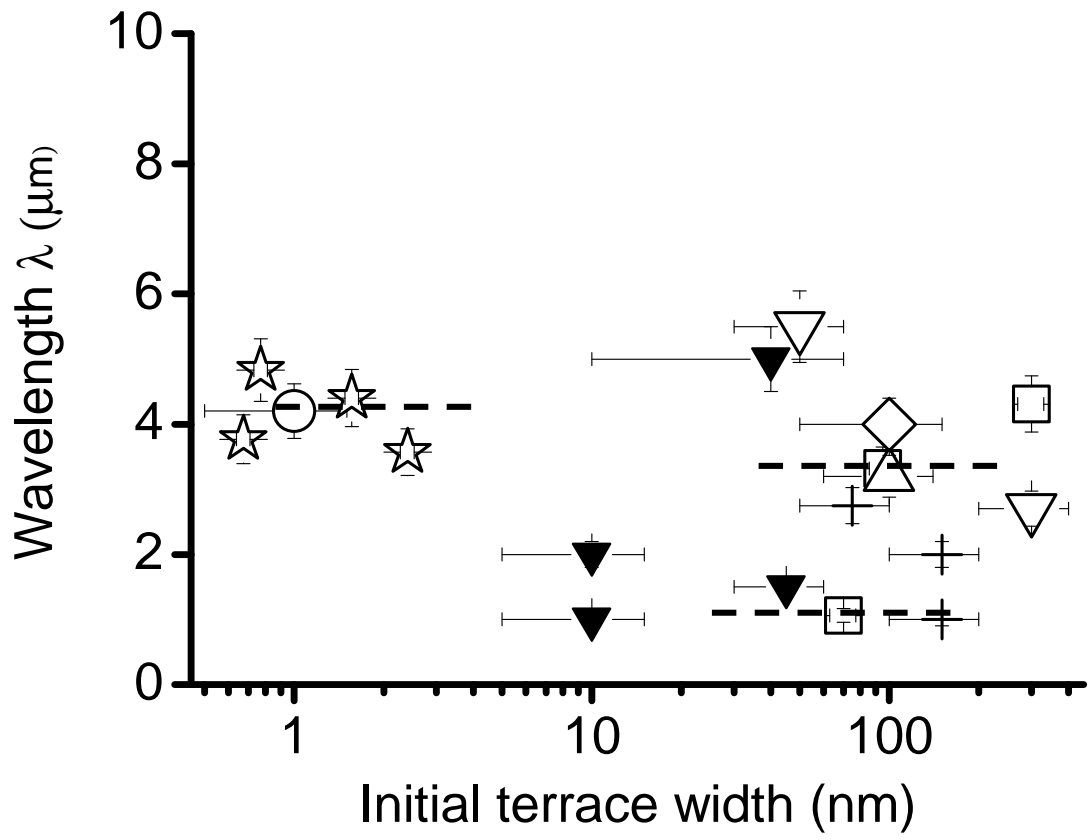

Figure 8 :

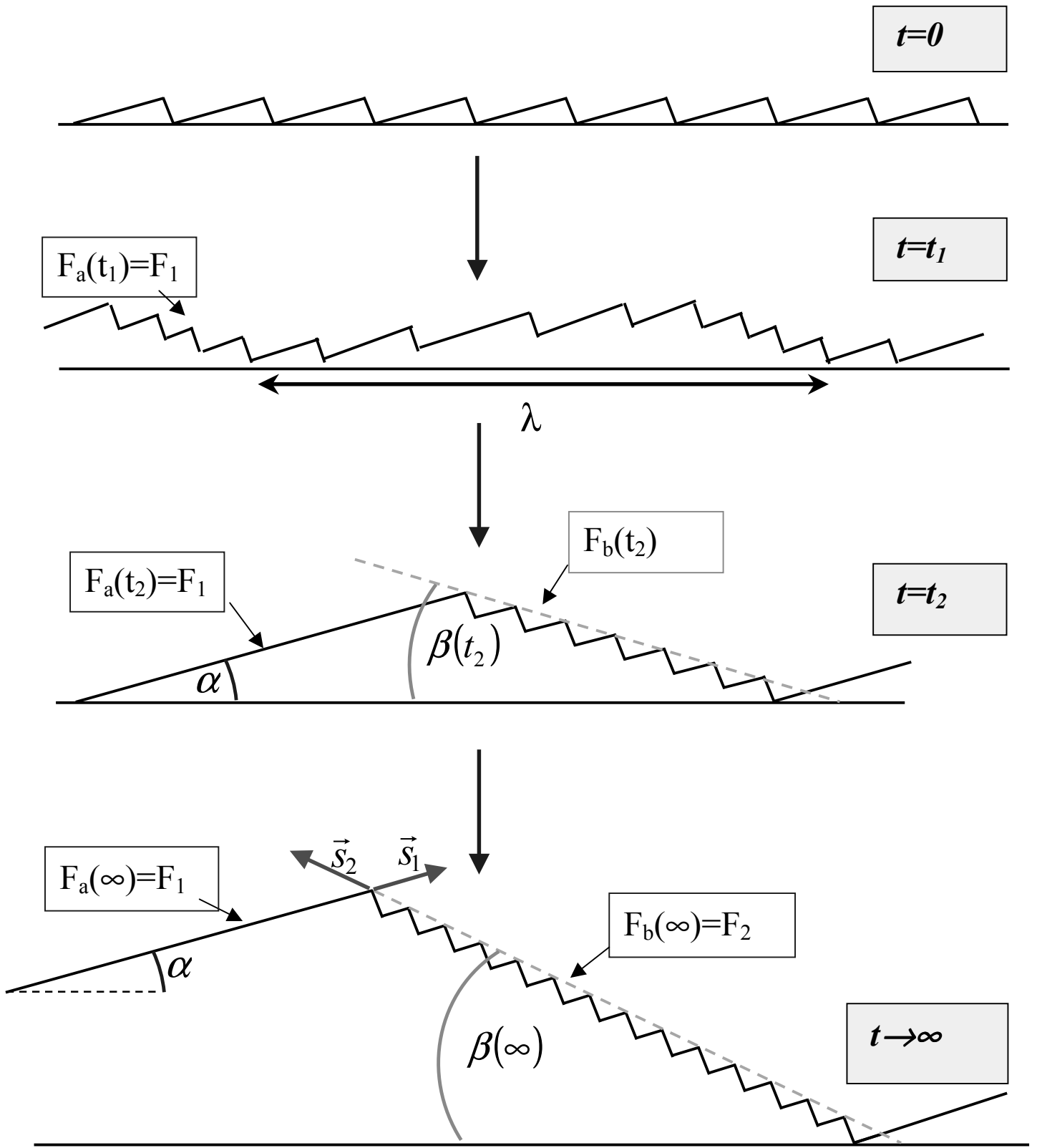

Fig 9:

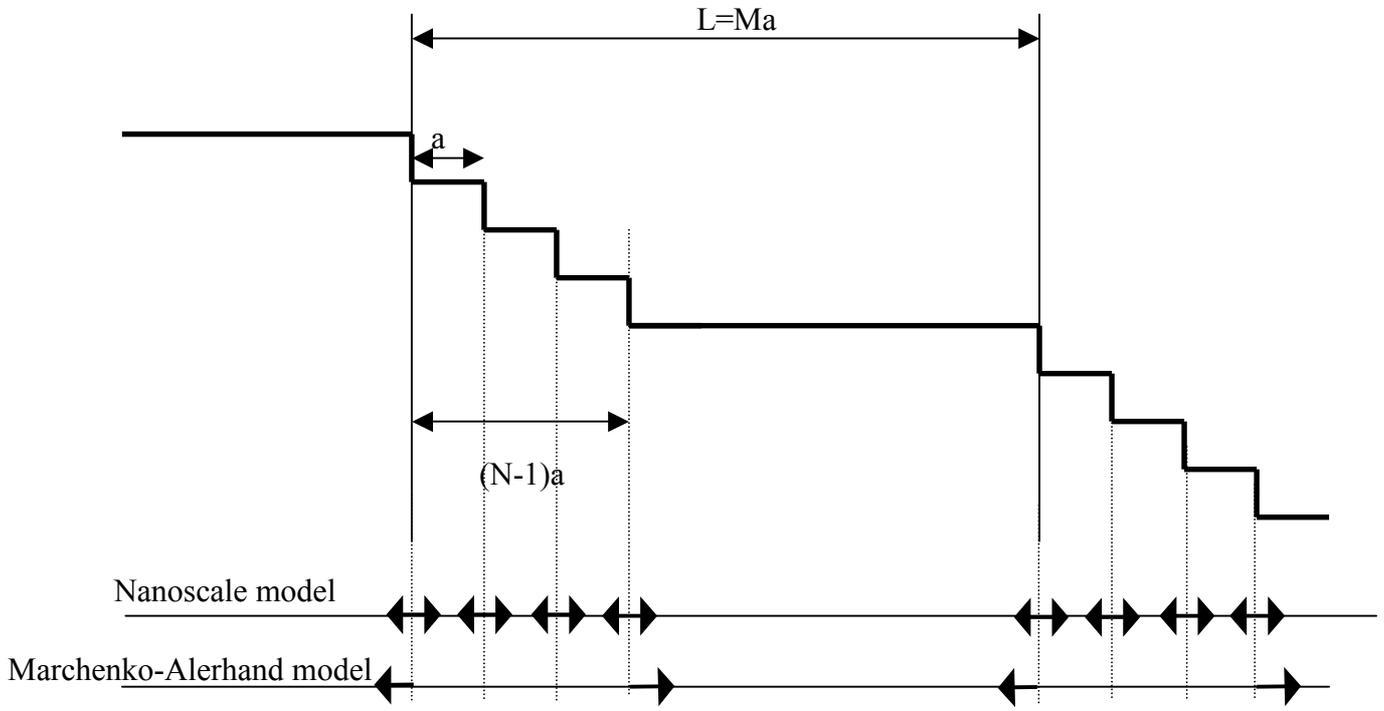

Fig 10 :

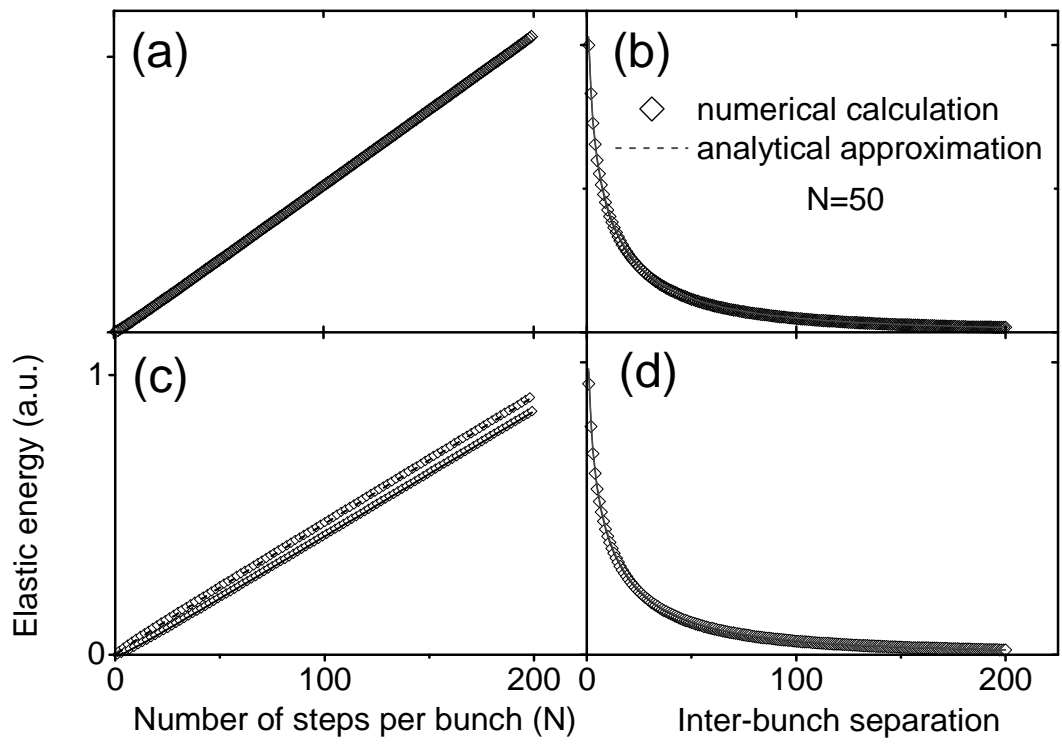

3939